\documentclass[preprint,12pt]{elsarticle}

\usepackage{graphicx}
\usepackage{amssymb}
\usepackage{textcomp}
\usepackage{url}
\usepackage{subfig}
\usepackage{jabbrv}
\usepackage{color}

\usepackage[top=1in, bottom=1in, left = 1in, right = 1in]{geometry}

\biboptions{square, sort&compress}

\journal{Acta Materialia}

\begin{document}

\begin{frontmatter}

\title{Predicting the Morphologies of $\gamma'$ Precipitates in Cobalt-Based Superalloys}

\author[CHiMaD,ANL-2]{A. M. Jokisaari\corref{cor1}}
\ead{andrea.jokisaari@northwestern.edu}
\cortext[cor1]{Corresponding author}

\author[NU]{S. S. Naghavi}
\ead{naghavi.shahab@northwestern.edu}

\author[CHiMaD,NU]{C. Wolverton}
\ead{c-wolverton@northwestern.edu}

\author[CHiMaD,NU]{P. W. Voorhees}
\ead{p-voorhees@northwestern.edu}

\author[NAISE,ANL-1]{O. G. Heinonen}
\ead{heinonen@anl.gov}

\address[CHiMaD]{Center for Hierarchical Materials Design, Northwestern University, 2205 Tech Drive, Suite 1160, Evanston, IL, 60208, USA}
\address[NU]{Department of Materials Science and Engineering, Northwestern University, 2220 Campus Drive, Evanston, IL 60208, USA}
\address[NAISE]{Northwestern-Argonne Institute of Science and Engineering, 2205 Tech Drive, Suite 1160, Evanston, Illinois 60208, USA}
\address[ANL-1]{Materials Science Division, Argonne National Laboratory, 9700 South Cass Avenue, Lemont, IL 60439, USA}
\address[ANL-2]{Physical Sciences and Engineering Directorate, Argonne National Laboratory, 9700 South Cass Avenue, Lemont, IL 60439, USA}

\begin{abstract}

Cobalt-based alloys with $\gamma$/$\gamma'$ microstructures have the potential to become the next generation of superalloys, but alloy compositions and processing steps must be optimized to improve coarsening, creep, and rafting behavior. While these behaviors are different than in nickel-based superalloys, alloy development can be accelerated by understanding the thermodynamic factors influencing microstructure evolution.  In this work, we develop a phase field model informed by first-principles density functional theory and experimental data to predict the equilibrium shapes of Co-Al-W $\gamma'$ precipitates.  Three-dimensional simulations of single and multiple precipitates are performed to understand the effect of elastic and interfacial energy on coarsened and rafted microstructures; the elastic energy is dependent on the elastic stiffnesses, misfit strain, precipitate size, applied stress, and precipitate spatial distribution.  We observe characteristic microstructures dependent on the type of applied stress that have the same $\gamma'$ morphology and orientation seen in experiments, indicating that the elastic stresses arising from coherent $\gamma$/$\gamma'$ interfaces are important for morphological evolution during creep.  The results also indicate that the narrow $\gamma$ channels between $\gamma'$ precipitates are energetically favored, and provide an explanation for the experimentally observed directional coarsening that occurs without any applied stress. 

\end{abstract}


\end{frontmatter}

\section{Introduction}
\label{sec:Introduction}

Superalloys are a broad class of alloys that exhibit high strength and oxidation resistance for high-temperature applications  \cite{donachie2002superalloys}. Nickel-based superalloys exhibit a $\gamma/\gamma'$ microstructure, in which ordered $\gamma'$ phase inclusions are embedded in a $\gamma$-phase matrix. Both the  $\gamma$ and  $\gamma'$ phases have fcc crystal structures, but the $\gamma$ phase is a disordered solid solution, while the $\gamma'$ phase is an ordered L$1_2$ structure. This $\gamma/\gamma'$ microstructure leads to high-temperature creep resistance because of the large interfacial area between the $\gamma$ and $\gamma'$ phases, which blocks the motion of dislocations \cite{pollock2006nickel}. Because of this, nickel-based superalloys are used in high-stress high-temperature applications such as gas turbine engines.

The efficiency of gas turbine engines generally increases with the operating temperature of the engine, but further increases in operating temperature require the development of new types of superalloys. A promising class of candidates for next-generation superalloys are cobalt-based superalloys with $\gamma/\gamma'$ microstructures. Because the melting temperature, oxidation, and wear resistance of cobalt is higher than that of nickel, the maximum operating temperature of $\gamma/\gamma'$ cobalt-based superalloys may be greater than that of nickel-based superalloys, which would improve turbine efficiency and lead to substantial fuel cost savings. While $\gamma$-phase cobalt-based superalloys have been used for decades in static, low stress service conditions \cite{donachie2002superalloys}, their poor creep resistance has precluded their use in high-stress applications. However, a $\gamma/\gamma'$ microstructure in Co-Al-W was reported in 2006 \cite{sato2006cobalt}, spurring a new wave of research into cobalt-based superalloys.  

To supplant nickel-based superalloys, $\gamma/\gamma'$ cobalt-based superalloys must exhibit better mechanical properties at higher operating temperatures. The microstructure is not static during service, however, meaning that creep tests spanning hundreds or thousands of hours must be performed to characterize new alloys.  Coarsening of $\gamma'$ occurs, in which some precipitates grow at the expense of others. When coarsening occurs under an applied stress and plate-like or rod-like precipitate shapes develop, the process is termed ``rafting'' \cite{nabarro1996rafting}. Coarsening and rafting affect mechanical properties because of the concomitant loss of interfacial area and change in precipitate shape \cite{kamaraj2003rafting}. Coarsening, creep, and rafting behaviors are different in cobalt-based $\gamma/\gamma'$ superalloys than in nickel-based superalloys: both creep and coarsening rates and precipitate shape and alignment under applied stress are different.  These differences arise from a complex interplay of the variations in interfacial energies, misfit strains, and elastic stiffnesses from system to system \cite{bauer2012creep,coakley2017rafting,suzuki2015l12}. Because these processes occur over long time scales, an understanding of the thermodynamic driving force affecting microstructural evolution should accelerate alloy development. 

Developing a new superalloy may take a decade or more, as careful control of composition and processing steps are required to optimize the material properties. Therefore, there is significant motivation to utilize integrated computational materials engineering (ICME) to accelerate the development cycle, a method that has already engendered several commercial successes for other metal alloys 
\cite{kuehmann2009computational}.  In ICME, materials models and experimental data at multiple length scales are linked together.  Mesoscale modeling, which examines materials at the nanometer to micron length scale, can improve understanding of superalloy microstructural evolution. Mesoscale studies have investigated both the equilibrium shapes of individual $\gamma'$ precipitates \cite{thompson1994equilibrium,schmidt1997equilibrium, jou1997microstructural, leo1998diffuse, mueller19983d,schmidt1998effect, thompson1999equilibrium,mueller19993d,jog2000symmetry,leo2000microstructural,kolling2003influence,li2003microstructure,li2004two,zhao2013effects,zhao2015equilibrium} and the coarsening behavior of multi-precipitate systems \cite{akaiwa2001large,zhu2004three,wang2008coarsening,tsukada2009phase,zhou2010large,boussinot2010phase,kundin2012phase,zhou2014computer,cottura2015role,mushongera2015effect,mushongera2015phase,tsukada2017phase}, with the former generally studied via sharp-interface models and the latter primarily via phase field models, though both approaches have been taken to study both problems.  Studies on the equilibrium shapes of precipitates have been primarily 2D in nature \cite{thompson1994equilibrium,schmidt1997equilibrium,jou1997microstructural,leo1998diffuse,schmidt1998effect, jog2000symmetry,leo2000microstructural,kolling2003influence,zhao2013effects,zhao2015equilibrium}, while fewer 3D studies have been performed \cite{mueller19983d,mueller19993d,thompson1999equilibrium,li2003microstructure,li2004two}.  These works have studied either nickel-based superalloys or generalized cubic materials systems.  To our knowledge, no studies have been performed to quantitatively predict equilibrium precipitate shapes of cobalt-based superalloys, or to predict the shapes of multi-precipitate arrays under applied stress in three dimensions.

The theoretical focus of many prior mesoscale studies provides an important framework to study $\gamma$/$\gamma'$ cobalt-based superalloys.  It has been shown that the equilibrium shape of an elastically stressed precipitate with isotropic interfacial energy may not be cubic even though the crystal system of both the matrix and precipitate are cubic, a phenomenon termed ``a shape bifurcation'' \cite{johnson1984elastically}.  The $L'$ parameter \cite{li2004two},
\begin{equation}
L' \equiv \frac{\overline{g_{el}} \, l}{\Gamma},
\label{eq:Lprime}
\end{equation}
is helpful in analyzing the bifurcation behavior of a precipitate.  In Eq.~(\ref{eq:Lprime}), $\overline{g_{el}}$ is a characteristic dimensional strain energy density, $l$ is a characteristic length of the precipitate equal to the radius of a sphere of the same volume, and $\Gamma$ is the interfacial energy per unit area. Thus, $L'$ is a non-dimensional length that characterizes the ratio of a precipitate's elastic energy and interfacial energy. Differences in precipitate morphology at the same $L'$ value can thus be studied to understand the impact of crystal symmetries, elastic stiffness variations, and anisotropy of the interfacial energy and misfit strain.

The focus of this work is to understand the energetics that drive the equilibrium shapes of $\gamma'$ precipitates in a $\gamma$ matrix, both in stress-free conditions as well as under applied stress. Using input parameters from experiments or first-principle calculations, we predict the equilibrium shapes of $\gamma'$ precipitates in the Co-Al-W system to help understand the thermodynamic forces driving the system's coarsening and rafting behavior. For simplicity, we choose the Co-Al-W system, which has the fewest number of alloying elements and a relatively large body of published literature compared to other cobalt-based $\gamma$/$\gamma'$ superalloys; our results should be qualitatively applicable to systems with additions of minor alloying elements that do not change the sign of the misfit strain. To predict precipitate morphologies, we develop a phase field model that incorporates elastic energy and interfacial energy without directly including phase compositions. We find that precipitate morphologies compare very favorably to those found with sharp-interface approaches, with the added benefit of incorporating the diffuse nature of the $\gamma$/$\gamma'$ interface observed with atom-probe tomography \cite{meher2013coarsening,povstugar2014elemental}.  We study single and multiple precipitates over a range of sizes with and without applied stress.  Our results predict characteristic microstructural features depending on the applied stress that agree well with experimentally-observed morphologies and provide an explanation for the experimentally observed directional coarsening that occurs without applied stress.  Furthermore, our work provides insight into how variations (and uncertainty) in elastic moduli, interfacial energy, and $\gamma$/$\gamma'$ misfit strain affect the final precipitate morphology, indicating which parameters must be measured carefully and for which more uncertainty is acceptable.

\section{Model formulation}
\label{sec:Model-formulation}

Our formulation is based on the phase field approach, in which a phase field $\eta$ takes the value of 0 in the $\gamma$ phase, 1 in the  $\gamma'$ phase, and varies smoothly across interfaces; other phase-dependent properties are interpolated between phases using $\eta$. We choose a phase field formulation to study the Co-Al-W system for several reasons.  Foremost, experimental evidence indicates that $\gamma$/$\gamma'$ interface is diffuse \cite{meher2013coarsening,povstugar2014elemental},
and diffuse interfaces between the phases naturally develop with the phase field method. While sharp-interface models have generally been employed to study the equilibrium shapes of precipitates, they do not capture the nature of the diffuse interface, they present significant challenges to modeling particle merging and splitting, and three-dimensional elasticity formulations for sharp-interface models are challenging. In addition, Ref.\ \cite{leo1998diffuse} demonstrated that diffuse-interface formulations with elasticity are equivalent to sharp-interface formulations as the diffuse interface width approaches zero.  

\subsection{Free energy and dynamics}
\label{subsec:free-energy-and-dynamics}

\begin{figure}
\centering
\includegraphics[scale=1]{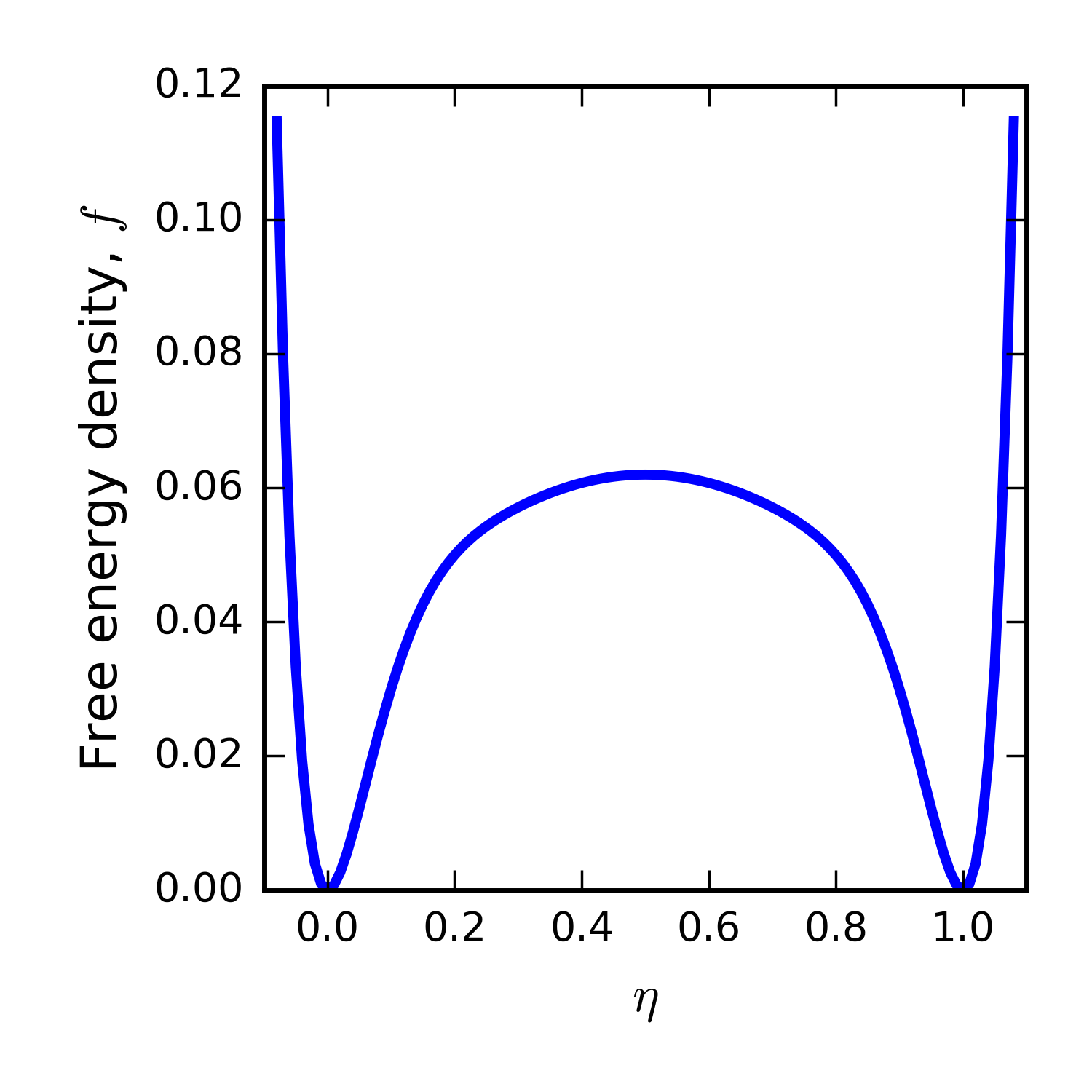}

\caption{The bulk free energy density, $f_{bulk}$, for $w=1$.  The energies of the equilibrium phases are zero, such that the only energy contributions to the system are interfacial and elastic energy. The energy wells are narrow and deep to prevent the actual value of $\eta$ in each phase from shifting significantly from its equilibrium value due to the presence of a curved interface or elastic strain; a large shift would introduce non-negligible error into the elastic energy calculation. \label{fig:free-energy}}
\end{figure}

For this work, we follow an approach similar to the diffuse-interface model presented in Ref.\ \cite{leo1998diffuse}.  We use a phase field free energy formulation that includes both the interfacial energy between the $\gamma$ and $\gamma'$ phases and linear elasticity; linear elasticity is a valid assumption given the small misfit strain between the phases. Anisotropic elastic stiffnesses and different elastic properties for each phase are incorporated.  As we are interested only in equilibrium precipitate morphologies, we do not explicitly incorporate chemical diffusion or chemical phase energies; chemistry is incorporated via the model parameterization for misfit strain, interfacial energy, and elastic stiffnesses. The free energy of the system, $F$, is expressed as 
\begin{equation}
F=\int_{V}\left(f_{bulk}\left(\eta\right) + \frac{\kappa}{2}|\nabla \eta|^{2} + f_{el}\left(\eta\right) \right)dV,
\label{eq:F_tot}
\end{equation}
where $f_{bulk}$ is formulated such the energies of the equilibrium phases are zero and it contributes only to the interfacial energy, $\kappa$ is the gradient energy coefficient, and $f_{el}$ is the local elastic energy density. To prevent the actual value of $\eta$ in each phase from shifting significantly from its equilibrium value due to the presence of a curved interface or elastic strain, an issue encountered in Ref.\ \cite{leo1998diffuse}, we choose $f_{bulk}$ to have a 10th-order polynomial form, 
\begin{equation}
f_{bulk}=w\sum_{j=0}^{10}a_j\eta^j,
\label{eq:f_chem}
\end{equation}
where $w$ controls the height of the energy barrier, so that the energy wells of the $\gamma$ and $\gamma'$ phases are deep (Fig. \ref{fig:free-energy}). This concern is discussed further in Section \ref{sub:ICS-validation}. The $f_{bulk}$ coefficients are given in Table \ref{tab:fbulk-params}, such that $f_{bulk}\left(0\right)=f_{bulk}\left(1\right)=f_{bulk}'\left(0\right)=f_{bulk}'\left(1\right)=0$ and the energy curve remains concave down between the two energy wells.

\begin{table}
\centering
\caption{\label{tab:fbulk-params}Parameterization of the $f_{bulk} = w\sum_{j=0}^{10}a_j\eta^j$ term. The large number of significant digits are necessary to ensure that the first derivative of $f_{bulk}$ is zero at $\eta=0$ and $\eta=1$.}
\begin{tabular}{c c c c}
\hline
$a_j$ & parameter value & $a_j$ & parameter value \\\hline
$a_0 = a_1$ & 0  & $a_6$ & 2444.046270 \\
$a_2$ & 8.072789087 & $a_7$ &  -3120.635139\\
$a_3$ & -81.24549382 & $a_8$ &  2506.663551\\
$a_4$ & 408.0297321 & $a_9$ & -1151.003178\\
$a_5$ & -1244.129167 & $a_{10}$ & 230.2006355\\
\hline
\end{tabular}

\end{table}
The elastic energy density is given as 
\begin{equation}
f_{el}\left(\eta\right)=\frac{1}{2}\sigma_{ij} \epsilon_{ij}^{el},
\label{eq:f_el}
\end{equation}
where $\sigma_{ij}=C_{ijkl}\left(\eta\right) \epsilon_{ij}^{el}$ is the elastic stress, $\epsilon_{ij}^{el}$ is the elastic strain, and $C_{ijkl}\left(\eta\right)$ is the elastic stiffness tensor (the Einstein summation convention is used).  Because the lattice parameters of the two phases are different, the elastic strain differs from the total strain, $\epsilon_{ij}^{total}$, as \cite{eshelby1957determination}
\begin{equation}
\epsilon_{ij}^{el}=\epsilon_{ij}^{total}-\epsilon_{ij}^{0}\left(\eta\right),
\label{eq:elastic_strain}
\end{equation}
where $\epsilon_{ij}^{0}$ is the local stress-free strain.  The total strain is calculated from the displacements, $u_i$, as \cite{eshelby1957determination}
\begin{equation}
\epsilon_{ij}^{total}=\frac{1}{2}\left[\frac{\partial u_i}{\partial x_j}+\frac{\partial u_j}{\partial x_i}\right],
\label{eq:total_strain}
\end{equation}
and the stress-free strain is calculated as 
\begin{equation}
\epsilon_{ij}^0\left(\eta\right)= \epsilon_{ij}^T \, h\left(\eta\right),
\label{eq:misfit_strain}
\end{equation}
where $\epsilon_{ij}^{T}$ is the crystallographic misfit strain tensor between the $\gamma$ and $\gamma'$ phases defined with respect to the $\gamma$ phase, and $h\left(\eta\right)=\eta^3\left(6\eta^2-15\eta+10\right)$, which ensures that $h\left( 0 \right)=h'\left( 0 \right)=h'\left( 1 \right)=0$ and $h\left( 1 \right)=1$ \cite{leo1998diffuse}.  To incorporate the phase dependence of the elastic stiffness, $C_{ijkl}\left(\eta\right)$ is given as
\begin{equation}
C_{ijkl}\left(\eta\right)= C_{ijkl}^{\gamma}\left[1-h\left(\eta\right)\right]+C_{ijkl}^{\gamma'} \, h\left(\eta\right),
\label{eq:stiffness}
\end{equation}
where $C_{ijkl}^{\gamma}$ and $C_{ijkl}^{\gamma'}$ are the stiffness tensors of the $\gamma$ and $\gamma'$ phases, respectively, and the stiffness is interpolated smoothly from one phase to the other across the diffuse interface.

In order to allow the precipitate shapes to equilibrate, we employ the the Cahn-Hilliard equation to perform fictive time evolution, as done in Ref.\ \cite{leo1998diffuse}. The evolution of $\eta$ is given as 
\begin{equation}
\frac{\partial\eta}{\partial t}=\nabla\cdot\left[M\nabla\left\{ \frac{\delta F}{\delta\eta}\right\} \right],
\label{eq:CH}
\end{equation}
where $M$ is the mobility, which we have flexibility in choosing as we are only interested in the final state of the system, and 
\begin{equation}
\frac{\delta F}{\delta\eta}=\frac{\partial f_{chem}}{\partial\eta}+\frac{\partial f_{elastic}}{\partial\eta}-\kappa\nabla^{2}\eta.\label{eq:dFdn}
\end{equation}
Note that the left-hand side of Eq.\ \ref{eq:dFdn} is the chemical potential, denoted as $\mu$, which we refer to in in Section \ref{sub:num-methods}.

\subsection{Model parameterization}
\label{subsec:model-parameterization}

In order to obtain quantitatively predictive results for the equilibrium shapes of Co-Al-W $\gamma'$ precipitates,  experimental or atomistic modeling input for the values of the interfacial energy, interface thickness, misfit strain, and elastic constants are needed; these parameters have been either measured or are calculated for this work.  We do increase the interface width in certain simulations in order to reduce the computational requirements, but we confirm that the morphology is not affected (the details are discussed further in Section \ref{sub:ICS-validation}). In addition, superalloy microstructures  evolve at high temperature (e.g., 1073 K -- 1273 K), but oftentimes measurements of material properties are performed at room temperature, which introduces a potential source of error into the model. In order to address this, we investigate two parameter sets, one for 300 K and the other for 1173 K; we discuss how these parameter sets are obtained in the following paragraphs.

First, we discuss the parameters for the $\gamma$/$\gamma'$ interface.  Using density functional theory (DFT), we calculate the interfacial energy at 0 K as 98 mJ/m$^2$ for the $\{100\}$ planes (see Sec.\ \ref{sub:num-methods} for details of the DFT calculations). We assume an isotropic interfacial energy based on the spherical shape at micron sizes of nickel- and cobalt-based superalloy precipitates with approximately zero misfit \cite{conley1989effect,ricks1983growth,meher2016solute}. For simplicity, we neglect any effect that temperature may have on the interfacial energy, but note that thermal effects may be significant. In addition, the $\gamma$/$\gamma'$ interface is diffuse, with a width of approximately 3 to 5 nm as measured by atom probe tomography \cite{meher2013coarsening,povstugar2014elemental}.  We choose an interface width of 5 nm for $0.05<\eta<0.95$ when the initial precipitate size is 100 nm or smaller; otherwise we generally use a larger interface width, targeting a 20:1 ratio of the precipitate diameter to the interface width to ensure a very large ratio of the bulk to interfacial regions.  We calculate $\kappa$ and $w$ numerically for each interface width to maintain a fixed interfacial energy (Table \ref{tab:interface-info}). 

\begin{table}
\centering
\caption{\label{tab:interface-info}Phase field parameters for an interfacial energy of 98 $\textrm{mJ/m}^2$.}
\begin{tabular}{c c c}
\hline
Interface width (nm) & $\kappa \, \left(\textrm{aJ/nm}\right)$ & $w \, \left(\textrm{aJ/nm}^{3}\right)$ \rule{0pt}{2.6ex} \rule[-0.9ex]{0pt}{0pt} \\\hline
5 & 0.58 & 0.195 \rule{0pt}{2.6ex}  \\
7.5 & 0.85 & 0.133 \\
10 & 1.14 & 0.100 \\
17.5 & 2.00 & 0.057 \\
25 & 2.85 & 0.040 \\
50 & 5.40 & 0.021 \\
\hline
\end{tabular}

\end{table}

Next, we discuss the parameters for the elastic behavior of the system, for which temperature dependence is incorporated.  The misfit strains of Co-Al-W or Co-Al-W-X alloys have been reported, primarily at room temperature \cite{sato2006cobalt,shinagawa2008phase,tanaka2012creep,pyczak2013plastic,yan2014alloying, zenk2014mechanical,pyczak2015effect}, with a limited number of high temperature measurements \cite{tanaka2012creep,pyczak2013plastic,yan2014alloying,pyczak2015effect}.  A range of misfit strain values have been reported at room temperature, but as the original value of 0.53\% published by Sato et al.\ \cite{sato2006cobalt} is in the middle of the range, we choose that, setting $\epsilon_{11}^{T}=\epsilon_{22}^{T}=\epsilon_{33}^{T}=0.53\%$.  The misfit reported at elevated temperature varies significantly, perhaps due to the addition of minor alloying elements, so we choose  $\epsilon_{11}^{T}=\epsilon_{22}^{T}=\epsilon_{33}^{T}=0.1\%$ \cite{pyczak2013plastic} as a lower bound at 1173 K.  We note that some error may be introduced by the misfit strains, because the values reported in the literature are calculated with constrained lattice parameters measured from two-phase material. Constrained lattice parameters may differ from their single-phase (unconstrained) values due to the presence of elastic stress. 

Because the $\gamma$ and $\gamma'$ phases are cubic, each phase has three independent elastic stiffness tensor values: $C_{1111}$, $C_{1122}$, and $C_{1212}$ (denoted in Voigt notation for a cubic system as $C_{11}$, $C_{12}$, and $C_{44}$, respectively).  The $C_{ijkl}^{\gamma}$ and $C_{ijkl}^{\gamma'}$ values at 300 K and 1173 K used in this work are given in Table \ref{tab:elastic-info}.  First, we discuss the $\gamma'$ phase.  The $C_{ijkl}^{\gamma'}$ values have been measured at 5 K \cite{tanaka2007single}, but not at elevated temperatures. However, they have been calculated via DFT at 0 K, 300 K and 1200 K \cite{xu2013first}. We also calculate the values at 0 K using DFT, which compare well with Refs.\ \cite{xu2013first,jiang2008first}.  The measured values at 5 K are in agreement with the DFT-calculated values at 0 K, so we use the values calculated in Ref.\ \cite{xu2013first} at 300 K and perform a linear interpolation to estimate the values at 1173 K.  

Conversely, the $C_{ijkl}^{\gamma}$ values have not been measured or calculated.  Therefore, we use a rule of mixtures to estimate the elastic constants at 300 K and 1173 K. To do so, we calculate the average composition of the $\gamma$ phase from values reported in Refs.\ \cite{kobayashi2009determination,lass2014gamma} and utilize experimentally measured single-crystal stiffnesses for fcc Co \cite{gump1999elastic,strauss1996martensitic}, fcc Al \cite{sutton1953variation,vallin1964elastic}, and bcc W \cite{featherston1963elastic,bolef1962elastic}. We again use the DFT approach to build confidence in our estimates.  First, we compute the elastic constants of fcc Co at 0 K, which compare well with the values calculated in Refs.\ \cite{guo2000gradient,pun2012embedded}.  Then, we calculate the elastic constants of Co-3.125\%Al, Co-3.125\%W, Co-6.25\%Al, Co-6.25\%W, and Co-3.125\%Al-3.125\%W at 0 K (Table \ref{tab:elastic-info}). Comparison of these DFT-calculated values with those found using the rule of mixtures indicates the error to be within $\pm$10\%, supporting our choice to use the values found by the rule-of-mixtures at 300 K and 1173 K.

\begin{table}
\centering
\caption{\label{tab:elastic-info}Calculated $C_{ijkl}$ values (in GPa) for different materials and compositions used in this work. All compositions are in atomic percent. $C_{ijkl}$ values of the $\gamma$ phase are for a composition of Co-9.1\% Al-5.3\% W.} 
\begin{tabular}{l l l l l}
\hline
Temperature and material & $C_{1111}$ & $C_{1122}$ & $C_{1212}$ & Ref.\ \rule{0pt}{2.6ex} \rule[-0.9ex]{0pt}{0pt} \\\hline
300 K, $\gamma'$ phase & 272 & 158 & 162 & \cite{xu2013first} \rule{0pt}{2.6ex}  \\
300 K, $\gamma$ phase & 229 & 165 & 97.3 \\[5pt]
1173 K, $\gamma'$ phase & 238 & 141 & 127 & \cite{xu2013first} \\
1173 K, $\gamma$ phase & 221 & 162 & 95.4 \\[5pt] 
0 K, fcc Co & 162 & 278 & 137 \\[5pt]
0 K, Co-3.125\% Al & 264 & 158 & 122 \\
0 K, Co-6.25\% Al & 277 & 159 & 148 \\[5pt]
0 K, Co-3.125\% W & 292 & 171 & 139 \\
0 K, Co-6.25\% W & 308 & 183 & 150 \\[5pt]
0 K, Co-3.125\% Al-3.125\% W & 276 & 170 & 140 \\

\hline
\end{tabular}
\end{table}

\section{Numerical methods and other simulation considerations}
\label{sec:Numerical-methods-sim-considerations}

\subsection{Numerical methods
\label{sub:num-methods}}

The simulations are performed with our MOOSE-based application, Hedgehog.  MOOSE \cite{gaston2014continuous,gaston2015physics} is an open-source finite element framework with several physics modules, including one for phase field modeling and one for tensor-based solid mechanics \cite{MooseModuleslink}.  To avoid prohibitively expensive fourth-order derivative operators in 3D, we split the fourth-order Cahn-Hilliard equation into two second-order equations \cite{elliott1989second,tonks2012object},  such that Eqs.\ \ref{eq:CH} and \ref{eq:dFdn} are solved separately with two different nonlinear variables.  The 3D-computational domains are cubic and are meshed with eight-node hexahedral elements.  Linear Lagrange shape functions are chosen for all nonlinear variables, and the system of nonlinear equations is solved using the preconditioned Jacobian-Free Newton-Krylov (PJFNK) method \cite{knoll2004jacobian}.  We apply the second backward differentiation formula (BDF2) \cite{iserles2009first} time integration scheme and use KSP preconditioning and LU factorization for sub-preconditioning.  

\begin{figure}
\centering
\subfloat[\label{fig:volfrac}]{\includegraphics[scale=1]{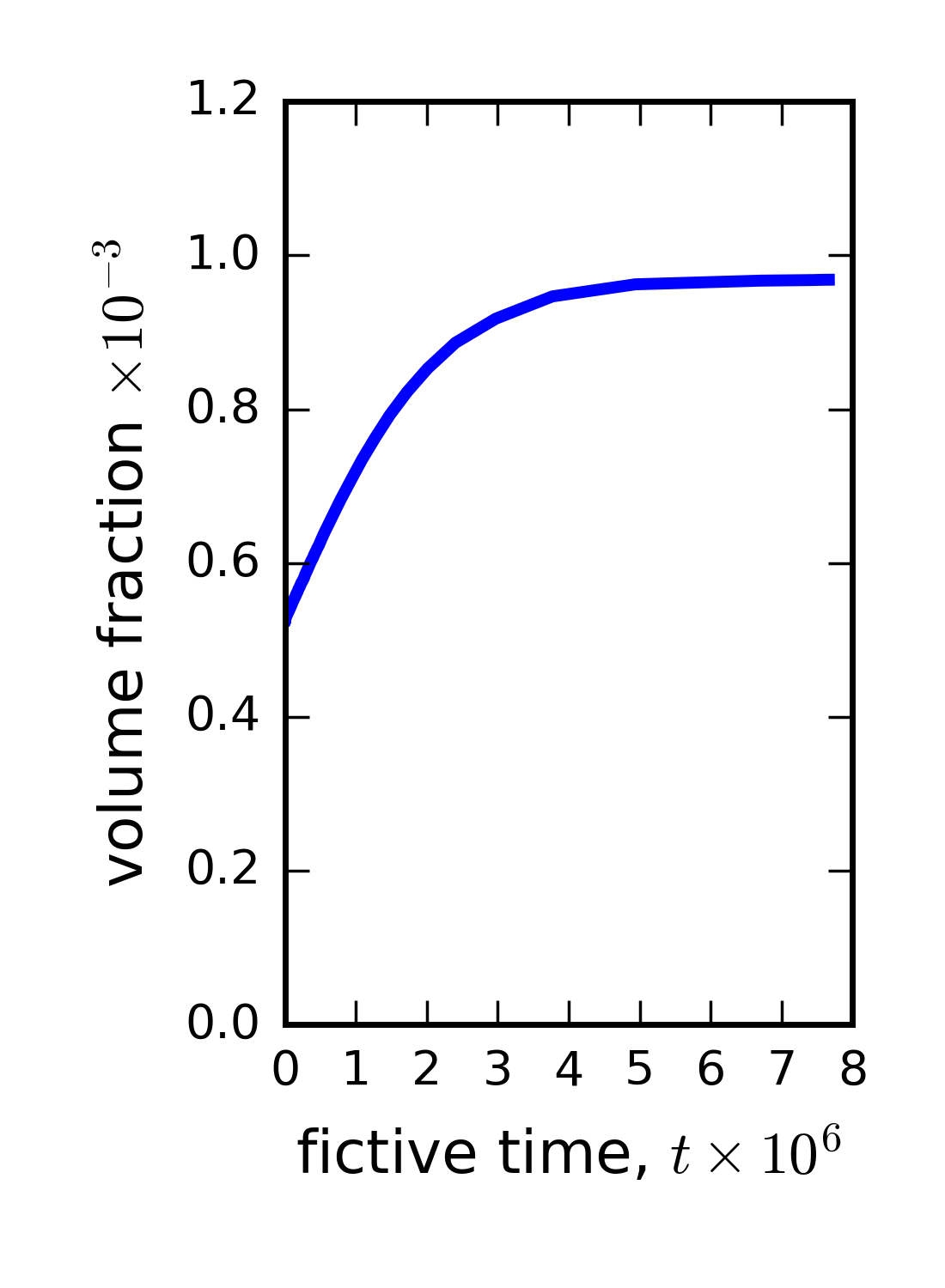}}
\subfloat[\label{fig:totalenergy}]{\includegraphics[scale=1]{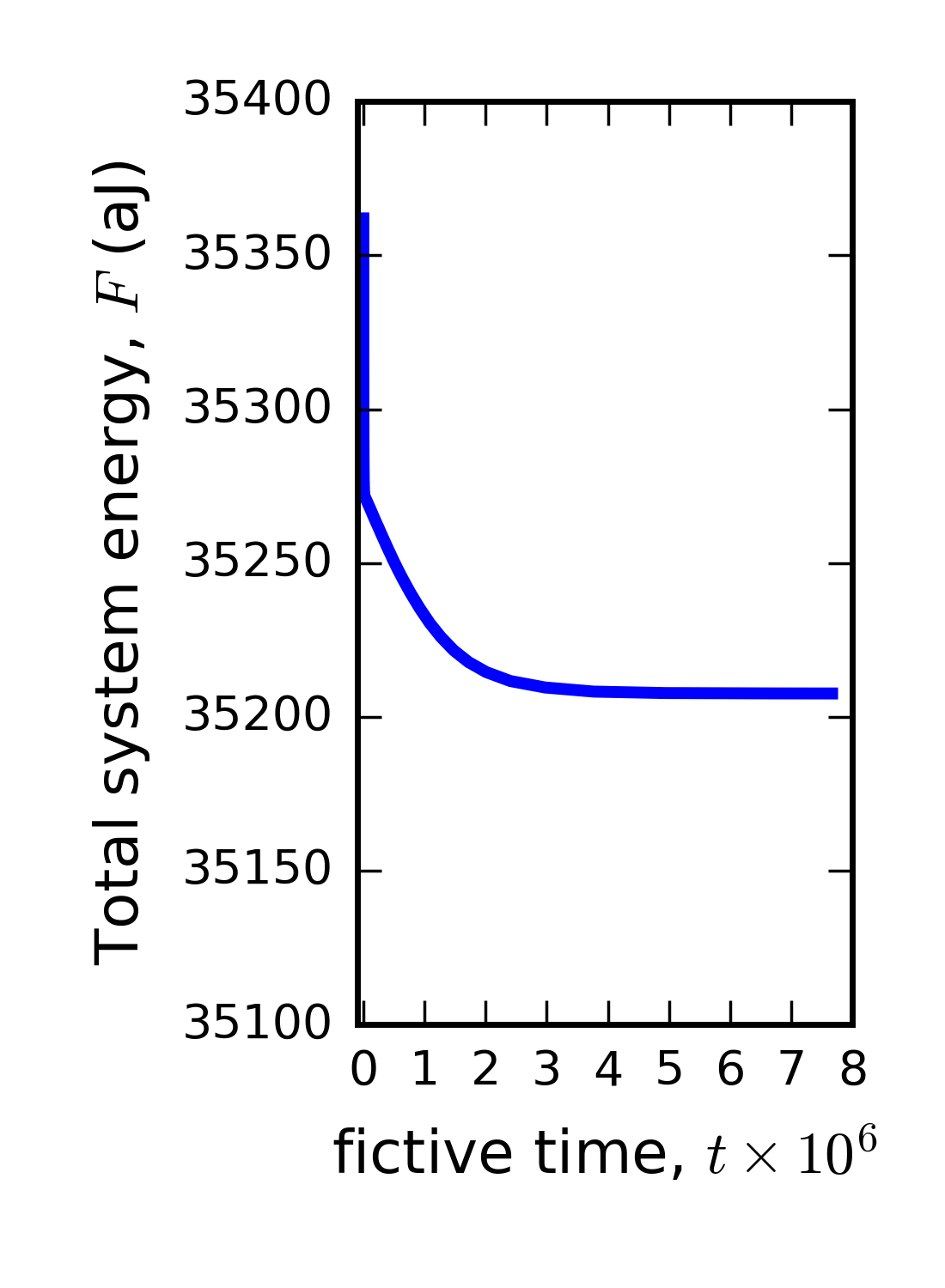}}
\subfloat[\label{fig:dt}]{\includegraphics[scale=1]{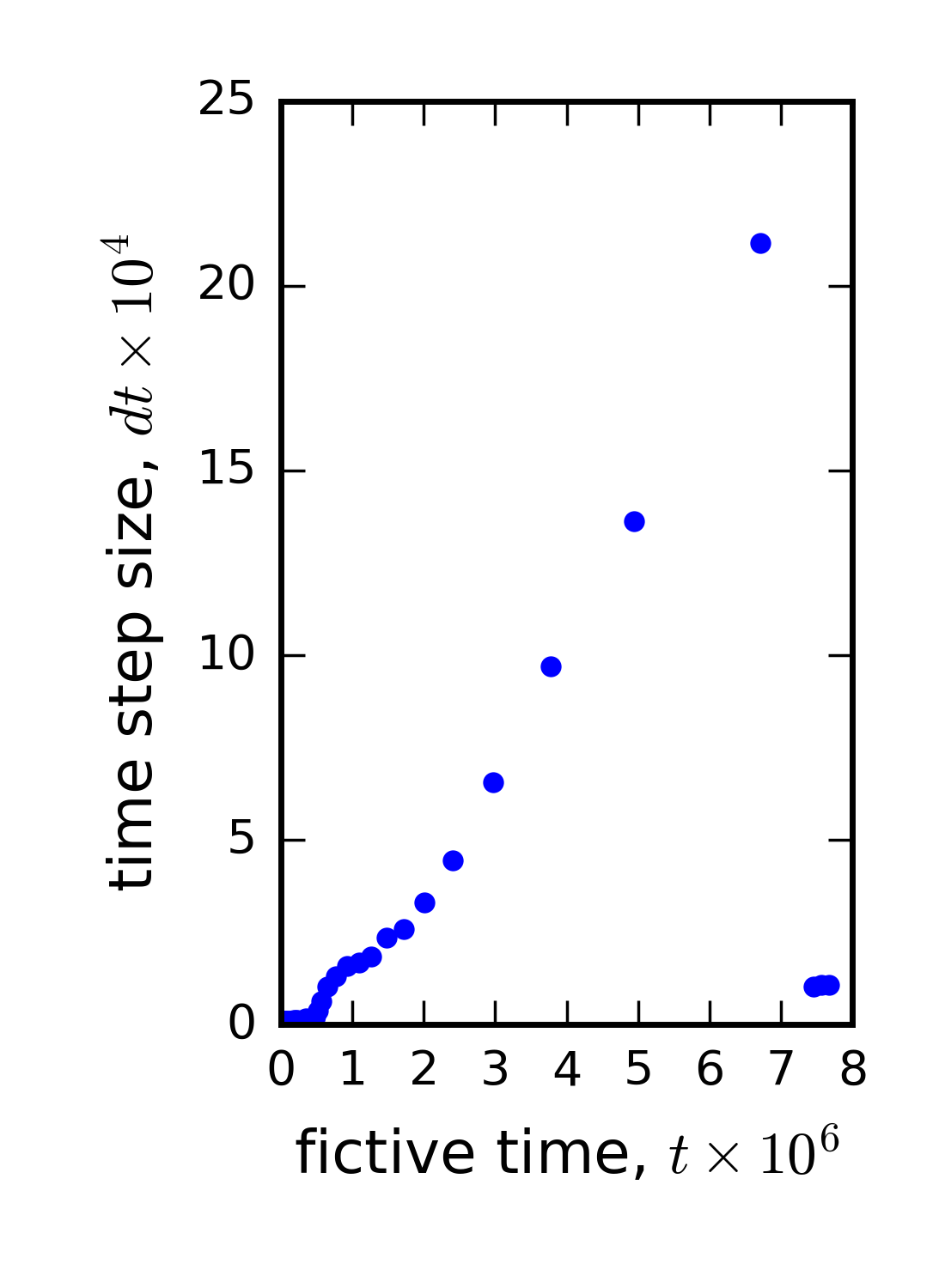}}

\caption{The relationship between the rate of precipitate evolution, the free energy, and the time step for the simulations in this work ($M=5$ for all simulations). The precipitate evolves rapidly early in the simulation, but as the  volume fraction and energy evolution slow, the time step ($dt$) size grows until the solver cannot converge the system to within numerical tolerances, at which point $dt$ collapses. a) Volume fraction of $\gamma'$, b) total system energy, and b) time step size. The data shown are from the simulation for a single precipitate with an initial diameter of 200 nm, no applied stress, and 300 K parameter set. \label{fig:system-evolution}}
\end{figure}

\begin{figure}
\centering
\subfloat[\label{fig:symmetry-domain}]{\includegraphics[scale=2]{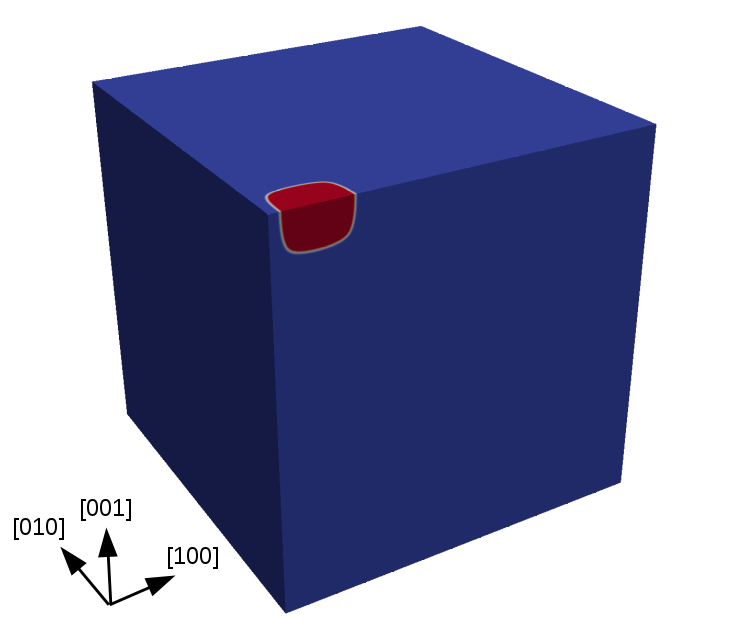}}
\subfloat[\label{fig:full-domain}]{\includegraphics[scale=2]{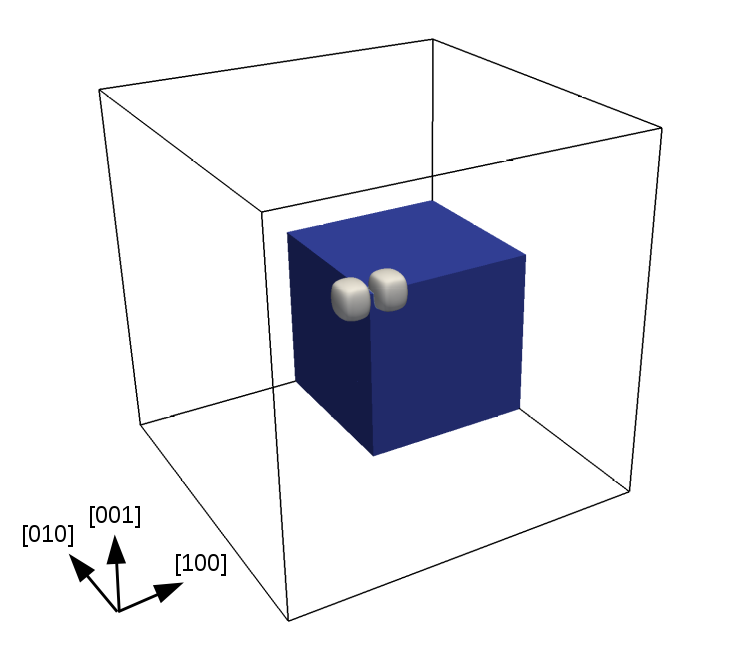}}

\caption{An illustration of the use of symmetry to reduce the computational domain size.  Here, a two-precipitate array has come to equilibrium.  a) The computational domain that is actually simulated, showing the location of the precipitate (in red). The simulated domain is one-eighth the size of the full domain. The mirror boundary planes cut across the center of the precipitate and through the narrow $\gamma$ channel separating it from the other, mirrored precipitate.  b) The full domain modeled by the one-eighth section (precipitate isosurfaces are shown in gray). \label{fig:symmetry-computational-domain}}
\end{figure}

We employ adaptive meshing, adaptive time stepping, and symmetry considerations to reduce the computational cost of the simulations. Gradient jump indicators \cite{kirk2006libmesh} for $\eta$, $u_1$, $u_2$, and $u_3$ are used to determine mesh adaptivity, and we ensure that the diffuse interface width spans at least five elements in all simulations.  The ``IterationAdaptive'' time stepper \cite{jokisaari2017benchmark} with a target of five nonlinear iterations per time step is chosen to govern the time step size. This time stepper greatly reduces the computational time for these simulations because it increases the time step as the driving force to evolve the system decreases. Eventually, the solver can no longer converge the system to within tolerances, at which point the time step collapses to its initial value, ending the simulation (Fig.\ \ref{fig:dt}). Examination of the simulation shows that $\mu$ is uniform by this time and that precipitate evolution has halted (Figs.\ \ref{fig:totalenergy} and \ref{fig:volfrac}).  Finally, we exploit symmetry to further reduce computational costs by a) utilizing the cubic symmetry of the materials and b) arranging the precipitates such that the mirror boundaries of the computational domain are along the symmetry axes of the arrangement.  One-eighth of the entire system is simulated and mirror boundary conditions are applied to the symmetry planes, such that the full system is modeled (the equilibrated two-particle system is shown in Fig.\ \ref{fig:symmetry-computational-domain} as an example). We check that this does not introduce errors, which we discuss further in Section \ref{sub:ICS-validation}. No-flux boundary conditions are applied to every boundary for $\eta$ and $\mu$, while the $u_1$, $u_2$, and $u_3$ displacements are pinned on the (100), (010), and (001) symmetry planes, respectively. For the external boundaries, natural boundary conditions are used for the displacements except when a stress is applied.

The DFT calculations are performed using the projector-augmented wave (PAW) method~\cite{PAW1,PAW2} as implemented in the Vienna Ab-initio Simulation Package (\texttt{VASP})~\cite{VASP1,VASP2}. VASP version 5.3 is used with the recommended PAW potentials for all elements \cite{PAWpotentials}. The exchange-correlation energy functional is described with the spin-polarized generalized gradient approximation (GGA) as parameterized by Perdew-Burke-Ernzerhof (PBE)~\cite{PBE}.  A kinetic cutoff energy of 348 eV is specified, and automatic k-point generation is used. To accurately evaluate the elastic properties, we set the VASP input flag PREC to ``PREC=high'', which specifies a plane wave cutoff energy 1.3 times that of the maximum energy cutoff listed in the PAW potentials. All structures are fully relaxed with respect to lattice vectors and atomic positions by minimizing the absolute energy until the Hellmann-Feynman forces on all atoms are less than 1~meV\,{\AA}$^{-1}$.

To calculate the $\gamma/\gamma^{\prime}$ interfacial energy, we build a supercell containing 16 atomic layers with a total of 64 atoms ~\cite{Geng2005}: eight layers of fcc cobalt and eight layers of Co$_{3}$(Al,W) quasi-random structures  \cite{jiang2008first}. This supercell structure is acceptable because we choose to work with the pure Co-Al-W system, avoiding the added complexity presented by impurity segregation. The interfacial energy is calculated in two steps. First, the total energy of the $\gamma/\gamma^{\prime}$ supercell, $E_{\gamma/\gamma'}$, is calculated with a full relaxation with respect to lattice vectors and atomic positions ($a$ and $b$ are lattice vectors in the plane of the interface, and $c$ is normal to both, along the long axis of the supercell). Then, the total energies of the single $\gamma$ and $\gamma^{\prime}$ phases ($E_{\gamma}$ and $E_{\gamma'}$, respectively) are calculated by fixing $a$ and $b$ to the previous step and relaxing $c$. The interfacial energy, $E_{\sigma}$, is given by

\begin{eqnarray}
 E_{\sigma}= \frac{E_{\gamma/\gamma^{\prime}}-E_{\gamma}-E_{\gamma^{\prime}}}{2S} ,
\end{eqnarray} 
where $S$ is the interfacial area, $S\!=\!a\times\!b$.

To calculate the elastic constants $C_{1111}$, $C_{1122}$, and $C_{1212}$ for the $\gamma'$ phase, we calculate the bulk modulus, $B$, and energies of the strained lattice.  The lattice is strained by a range of compressive and tensile distortions for both the monoclinic and orthorhombic strain types, and we ensure that volume is conserved.  The change in energy of the strained lattice versus the unstrained lattice, $\Delta E$, is fitted to the 4$^{th}$ order polynomial $\Delta E(x)=E_{0}+a_{2}x^{2}+a_{3}x^{3}+a_{4}x^{4}$, where $x$ is the strain.  To determine the values of $C_{1111}$, $C_{1122}$, and $C_{1212}$, $B$ and $a_{2}$ are used following Refs.\ \cite{Nye85} and \cite{Kart2010}.

\subsection{Initial conditions, computational domain size, validation
\label{sub:ICS-validation}}

In this section, additional simulation details such as initial conditions, computational domain size, and validation efforts are discussed. To study bifurcation behavior, we utilize both spherical and ellipsoidal initial precipitate shapes for a given initial precipitate volume \cite{thompson1994equilibrium,li2004two}. To maintain the same $l$ value (Eq.~\ref{eq:Lprime}) for an ellipsoid as for a sphere with radius $r$, we choose ellipsoid axes as $a_1=r$, $a_2=r/0.8$, and $a_3=0.8r$. In addition, the precipitate is embedded in a computational domain that is at least 10 times the size the initial precipitate radius to allow long-range elastic fields to decay.  

As noted in Section \ref{subsec:free-energy-and-dynamics}, the presence of a curved interface or elastic strain energy will shift the final $\eta$ value of each phase slightly.  Unlike sharp-interface approaches, which often conserve the total volume of precipitate, this model conserves the total integral of $\eta$ within the computational domain, such that the volume of the precipitate may change.  While this effect caused only a small amount of precipitate volume loss in Ref.\ \cite{leo1998diffuse} because the precipitate area filled a significant fraction of the computational domain, it is non-negligible in this work.  The volume of matrix within the computational domain is much larger than that of the precipitate, such that an initial precipitate can shrink completely away in the process of achieving the equilibrium value of $\eta$ in the matrix.  To avoid this, we set the initial value of $\eta$ in the matrix to be slightly greater than zero, e.g., 0.005.  In these simulations, the precipitate generally grows slightly (for example, see Fig.\ \ref{fig:volfrac}).

We perform several validation steps for our model and numerical methods. The work in Ref.\ \cite{leo1998diffuse} supports our choice of modeling approach; but we also compare to the 2D sharp-interface results in Ref.\ \cite{thompson1994equilibrium}.  We are not able to perfectly duplicate that parameter set, as only the anisotropy relationship between $C_{ijkl}$ values were given, not absolute values. However, we confirm that our model captures bifurcation behavior.  We test precipitates at $L=1$ and $L=10$ and find that four-fold symmetry occurs for the former size, while two-fold symmetry results at the latter.  In addition, we confirm that mirror boundary conditions do not alter the results by comparing the results of a 2D simulation performed with mirror boundary conditions to one which modeled the entire precipitate and matrix (without mirror boundaries). Further, we confirm that changing the interface width has a negligible effect on precipitate morphology by comparing the equilibrium shapes of a cuboidal precipitate with a size of 200 nm when simulated in 2D with an interface width of 5 nm and 10 nm.

\section{Results and discussion}
\label{Results-and-discussion}

In this section, we discuss the equilibrium shapes of Co-Al-W $\gamma'$ precipitates and the thermodynamic factors influencing the morphology. For this investigation, we perform multiple simulations of $\gamma'$ precipitates using both the 300~K and 1173~K parameter sets, and examine single precipitates of different sizes, multiple precipitates, and different applied stress states (no stress, uniaxial tension, and uniaxial compression).  Because elastically-driven bifurcation behavior may occur, we utilize both spherical and ellipsoidal initial shapes.  We discuss our results in the framework of the $L'$ parameter introduced in Section \ref{sec:Introduction}, which indicates the ratio of a precipitate's elastic to interfacial energy.  We also discuss characteristic microstructures that develop when stress is applied, as well as the effect of uncertainty in misfit strain, interfacial energy, and elastic stiffnesses on the results.  The input files, data, and code are available in the Materials Data Facility (DOI: XXX).

\subsection{Single precipitate
\label{sub:single-precip}}

\begin{figure}
\centering
\includegraphics[scale=1.75]{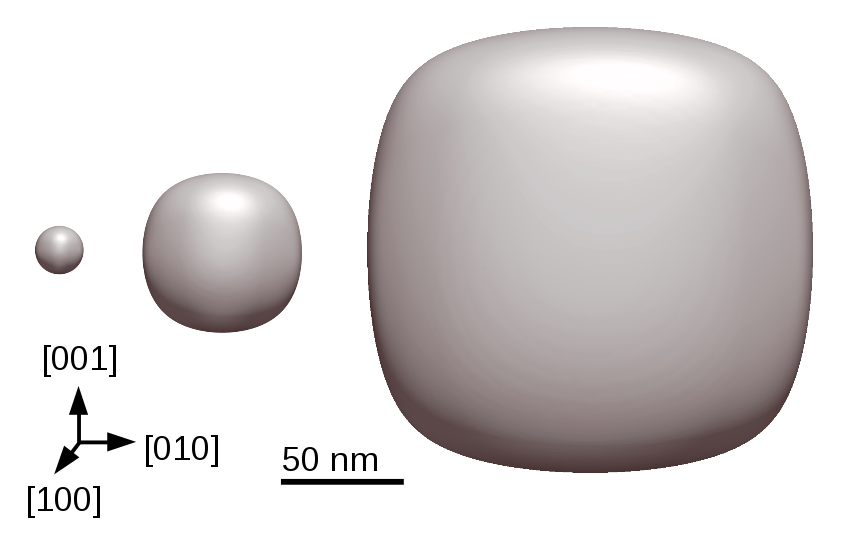}

\caption{The spherical to cuboidal transition for $\gamma'$ precipitates occurring over the 20 -- 180 nm size range.  Left) Spherical precipitate at a size of 20 nm, middle) rounded cuboidal precipitate at a size of 65 nm, right) cuboidal precipitate at a size of 180 nm.  \label{fig:sphere-to-cube}}
\end{figure}

\begin{figure}
\centering
\includegraphics[scale=1.75]{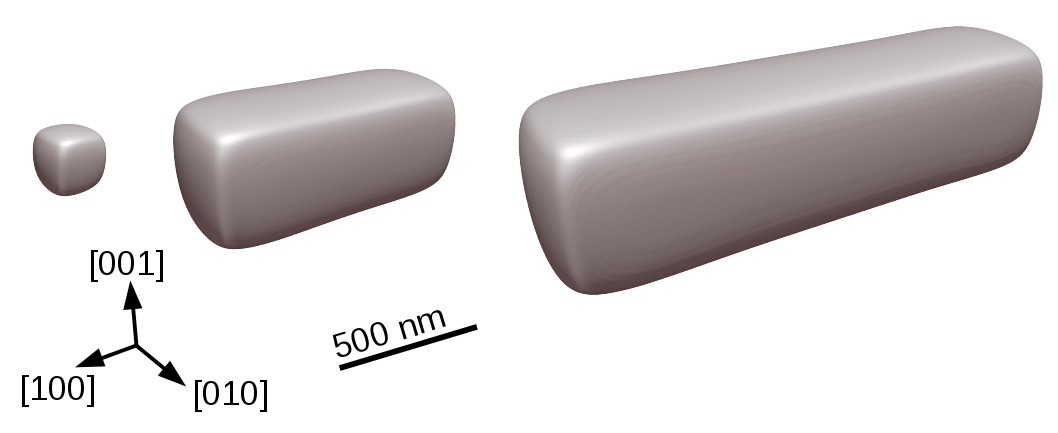}

\caption{The Co-Al-W $\gamma'$ precipitates exhibit two bifurcation behaviors. Left) Plate-like precipitates with very small axial ratios are observed over the range of 242 -- 272 nm; middle, right) rod-like precipitates are observed at larger sizes, and the axial ratio increases with increasing particle volume. \label{fig:bifurcation}}
\end{figure}

A fundamental understanding of Co-Al-W $\gamma'$ precipitate morphologies can be gained by studying single precipitates.  We start with the 300~K parameter set.  In this set, $C_{1111}^{\gamma'}=1.19\,C_{1111}^{\gamma}$, $C_{1122}^{\gamma'}=0.96\,C_{1122}^{\gamma}$, and $C_{1212}^{\gamma'}=1.66\,C_{1212}^{\gamma}$, indicating that the precipitate is harder than the matrix. A range of precipitate sizes are simulated, with the smallest having a diameter of 20~nm ($L'=0.28$).  As shown in Fig.\ \ref{fig:sphere-to-cube}, we observe a continuous transition from a spherical to a cuboidal shape as the  equilibrium size of the precipitates increases to 180~nm ($L'=2.8$), a result of the increasing ratio of elastic energy to interfacial energy within the precipitate.  At these sizes, the final precipitate exhibits cuboidal symmetry regardless of whether the initial precipitate shape is spherical or ellipsoidal. For larger precipitate sizes, different behavior is observed: precipitates with a spherical initial condition maintain cubic symmetry, but those with an ellipsoidal initial condition exhibit a lower order of symmetry (Fig.\ \ref{fig:bifurcation}).  This indicates that the cubic result is metastable \cite{thompson1994equilibrium,li2004two}, and the increase in total interfacial energy of the particle occurring with the change in symmetry is more than compensated for by the reduction in elastic energy.  We observe a plate-like shape for precipitates of 242~nm ($L'=3.9$) and 277~nm ($L'=4.4$) in size, indicating that bifurcation occurs at a precipitate size between 180 -- 242~nm ($2.8 < L' < 3.9$).  Both plate-like shapes have a very small axial anisotropy ratio, with the ratio of long:short axial lengths of 1.01.  At an even larger precipitate size, 319~nm ($L'=5.2$), the precipitate is rod-like, with a an axial ratio of ratio of 1.01, indicating a second shape transition between 277 -- 319~nm ($4.4 < L' < 5.2$).  The rod extends along one of the $\langle 100 \rangle$ directions. With increasing equilibrium precipitate volume, the precipitates remain rod-like and the axial ratio increases, reaching 3.55 for $L'=12.0$.  

While the axial ratio of the two plate-like shapes is very close to one, it is not a result of a numerical artifact.  Starting from the ellipsoidal initial condition, in which all three axial lengths of a precipitate are significantly different, the lengths of the precipitates' three axes evolve as the particles grow somewhat in volume.  First, all three axes shorten as the interfaces flatten under the influence of the cubic symmetry of the elastic stiffnesses and misfit. Axial lengths then start to increase; the time at which each axis starts to lengthen is different.  In this stage, $\mu$ is different for each of the precipitate/matrix interfaces normal to the axial directions, indicating that the shape is not equilibrated.  However, the total system energy is decreased much more rapidly by reducing the value of $\eta$ within the matrix by growing the precipitate.  Eventually, growth essentially halts, and the axial lengths adjust to their final values.  At the end of the simulation, the chemical potential is uniform throughout the entire computational domain to within five significant figures for one simulation, and within six for the other, indicating that the final shape is equilibrated. In addition, examination of the elastic and total system energies indicates that the simulations reached convergence tolerances.  Similar behavior is observed for the simulations in which ellipsoidal initial conditions are used that result in cuboidal and rod-like particles, indicating that the results reflect physical phenomena and not solver inaccuracies.

We first compare our results with those reported in Ref.\ \cite{li2004two} for single precipitates in a model Ni-superalloy system to understand the bifurcation behavior in the Co-Al-W system.  In Ref.\ \cite{li2004two}, four different precipitate/matrix stiffness ratios were studied, in which the $C_{ijkl}$ values of the precipitate were 0.9, 1, 1.1, and 1.5 times that of the matrix.  In that work, the critical $L'$ for bifurcation occurred somewhere within $3 < L' < 4$ for the system with homogeneous moduli, and by $L' \leq 5$ for every system except for the stiffest precipitate.  In addition, systems with softer precipitates exhibited only one shape bifurcation, but two shape bifurcations were observed with increasing size for the hardest precipitate system: first a plate-like precipitate shape with a small axial anisotropy ratio, followed by a rod-like precipitate shape. Furthermore, the axial ratio of the plate-like shape decreased as the precipitate became stiffer. 

In comparison, we observe what appears to be intermediate behavior between the different precipitate/matrix stiffness ratios studied in Ref.\ \cite{li2004two}.  We observe a critical $L'$ between 2.8 and 3.9, similar to the homogeneous moduli case in Ref.~\cite{li2004two}, yet we also observe plate-like and rod-like shape transitions, only seen for the stiffest precipitate in Ref.~\cite{li2004two}. Furthermore, the anisotropy for the plate-like particle in our work is even smaller than that observed for the stiffest precipitate system in Ref.~\cite{li2004two}. These differences are likely due to the more complex variation of $C_{1111}$, $C_{1122}$, and $C_{1212}$ with respect to each other for each individual phase as well as the their respective ratios in the $\gamma$ vs.\ the $\gamma'$ phases.  Our results show that  the ratio of $C_{1212}^{\gamma'}$ to $C_{1212}^{\gamma}$ is the controlling factor for the axial anisotropy ratio of plate-like shapes and the second bifurcation to rod-like shapes: our system had the largest ratio of $C_{1212}^{\gamma'}$ to $C_{1212}^{\gamma}$, the smallest axial anisotropy ratio, and the transition to a rod-like shape.

The simulated Co-Al-W $\gamma'$ morphologies also compare well with experimentally-observed microstructures. Rounded precipitates are observed at a size of 50 nm \cite{meher2013coarsening,tanaka2012creep}, while cuboidal precipitates are observed at a size of 100~nm (\cite{sato2006cobalt,meher2013coarsening}). These results also offer an explanation for the observed directional coarsening in the absence of applied stress \cite{sato2006cobalt,meher2013coarsening,pyczak2015effect}.  In directional coarsening, a precipitate elongates in one of the $\langle 100 \rangle$ directions and not all precipitates elongate in the same $\langle 100 \rangle$ direction. The rod-like equilibrium shape for precipitates larger than 320~nm indicate a thermodynamic driving force for this behavior: a small perturbation of a precipitate shape from cuboidal would be energetically favored to continue precipitate elongation in one of the $\langle 100 \rangle$ directions.

\begin{figure}\centering
\subfloat[\label{fig:1particle-compression}]{\includegraphics[scale=1.75]{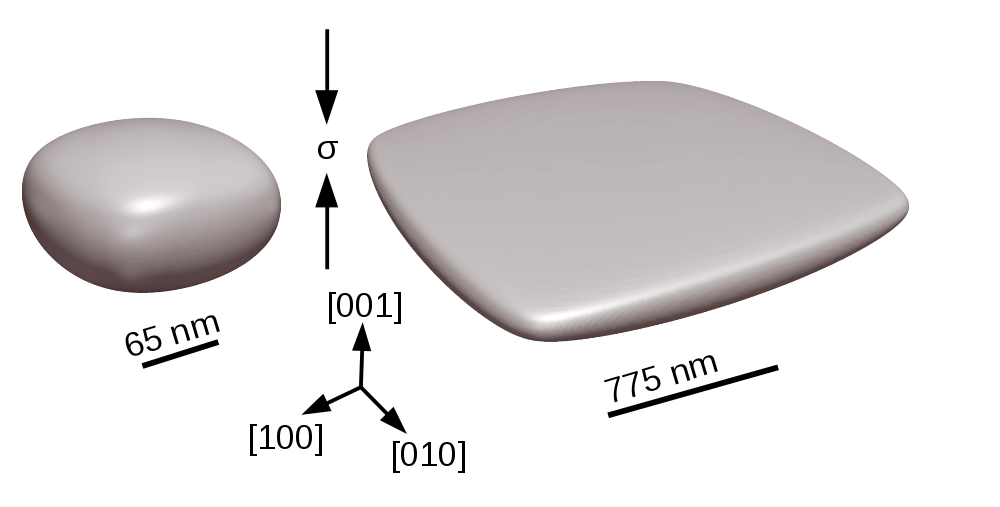}}
\subfloat[\label{fig:1particle-tension}]{\includegraphics[scale=1.75]{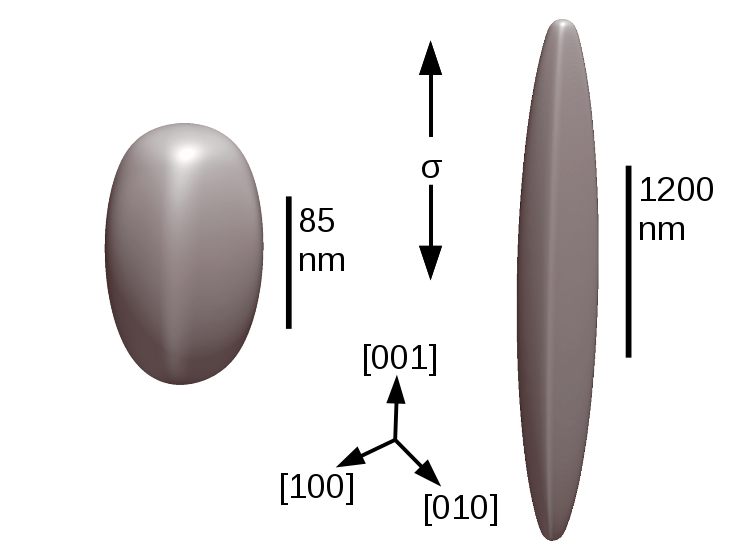}}

\caption{The morphology and orientation of a single precipitate is a function of its size and the direction of the applied stress (direction indicated by the arrows).  a) 400 MPa of compression results in a plate aligned normal to the stress,  b) 400 MPa of tension results in rods aligned parallel to the stress.  In both cases, the aspect ratio of the precipitates increases with increasing particle size. \label{fig:single-particle-stress-size}}
\end{figure}

Next, we study the behavior of precipitates under an applied stress. Precipitates of different sizes are allowed to evolve under the application of 400~MPa of uniaxial tension or uniaxial compression.  This stress value is chosen because it is similar to those used in creep testing \cite{pyczak2013plastic}.  We find that under tension, the precipitates are rod-like and oriented in the direction of the applied stress.  As seen in Fig.\ \ref{fig:1particle-tension}, the aspect ratio increases as the equilibrium size increases.  Under compression, the precipitates are plate-like and oriented normal to the direction of the applied stress, and the aspect ratio also increases with increasing precipitate size (Fig.\ \ref{fig:1particle-compression}).  The shape and orientation of the precipitates is a result of the positive misfit strain between the $\gamma$ and $\gamma'$ phases \cite{nabarro1996rafting} many nickel-based superalloys have a negative misfit strain between the phases and the shape and orientation of the precipitates is reversed \cite{nabarro1996rafting}.  Importantly, the shape and orientation of our simulated $\gamma'$ cobalt-based superalloy precipitates is also seen experimentally \cite{tanaka2012creep, bauer2012creep,coakley2017rafting}. This result indicates that elastic stresses arising from coherent interfaces are a significant driving force influencing precipitate microstructure evolution during creep.

\begin{figure}
\centering
\subfloat[\label{fig:1173K-smaller}]{\includegraphics[scale=1.75]{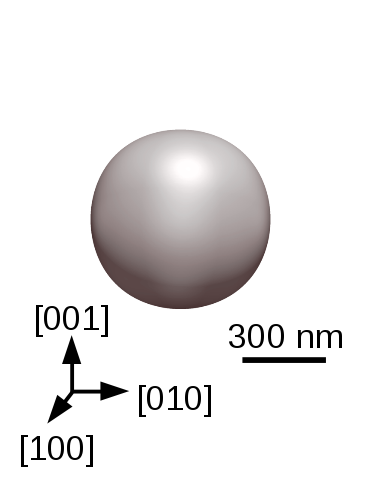}}
\subfloat[\label{fig:1173K-larger}]{\includegraphics[scale=1.75]{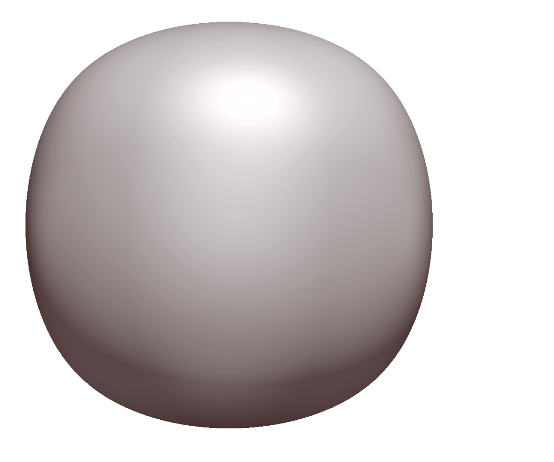}}
\subfloat[\label{fig:1173K-lowIntEnergy}]{\includegraphics[scale=1.75]{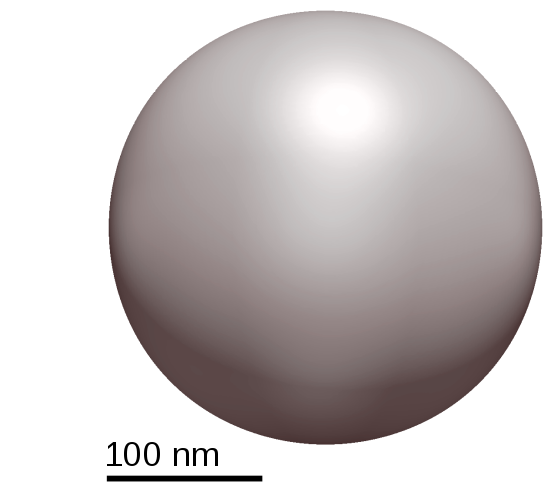}}

\caption{The morphology of a single precipitate as a function of size at 1173 K is far rounder than those seen in experimental microstructures, indicating that the ratio of elastic energy to interfacial energy is too small. a) At a diameter of 650 nm, the particle is very round, b) and becomes more cuboidal at a size of 1470 nm, while c) at a lower interfacial energy of 80 $\textrm{mJ/m}^2$, the particle is round at a diameter of 270 nm. \label{fig:1173K-particles}}
\end{figure}

The single-precipitate morphologies obtained with the 300~K parameter set agree well with experimental microstructures, so we now investigate the equilibrium shapes obtained with the 1173~K parameterization to explore how uncertainty impacts the results.  At 1173~K, $C_{1111}^{\gamma'}=1.0\,C_{1111}^{\gamma}$, $C_{1122}^{\gamma'}=0.87\,C_{1122}^{\gamma}$, and $C_{1212}^{\gamma'}=1.33\,C_{1212}^{\gamma}$, indicating that the $\gamma'$ phase is harder than the $\gamma$ phase, but significantly less so than in the 300 K parameterization.  In addition, we use a misfit strain of 0.1\% versus the 0.53\% at 300 K (Section \ref{subsec:model-parameterization}).  As before, we test a range of precipitate sizes without any applied stress, and find that precipitates remain spherical at diameters greater than 500 nm (Figs.\ \ref{fig:1173K-smaller}), which is in contradiction to experimental microstructures.  The spherical to cuboidal transition occurs gradually through a diameter 1475 nm (larger precipitate sizes were not tested), and bifurcation is not observed.  Given the possibility that entropic effects could reduce the interfacial energy at elevated temperature, we test a precipitate with a lower interfacial energy. We estimate a reduction of approximately 20 mJ/m$^2$ based on calculations for Ni-based superalloys  \cite{mao2012effects}, and use a value of 80 mJ/m$^2$.  For a precipitate with a diameter of 270 nm, similar in size to those observed experimentally, we find that the precipitate is still very round (Fig.\ \ref{fig:1173K-lowIntEnergy}).  

These results may be understood by examining the $L'$ values of the precipitates.  For the 1173~K system, $\overline{g_{el}}=9.45 \times 10^4\textrm{ mJ/m}^2$, while $\overline{g_{el}}=2.80 \times 10^6\textrm{ mJ/m}^2$ for the 300~K system, almost a thirty-fold difference.  As a result, a precipitate size of 1475~nm corresponds to  $L'=0.73$, far smaller than what we would reasonably anticipate as the bifurcation size.  Even with the lower interfacial energy of 80~mJ/m$^2$, $L'=0.16$ for the 270~nm diameter precipitate.  Because we are concerned with precipitate shape at an absolute size, we determine that the $\overline{g_{el}}$/$\Gamma$ ratio is much too small with this parameterization. 

While the parameterization at 300~K produces precipitate morphologies that agree well with experimental microstructures, the parameterization at 1173 K does not.  In the former case, the influence of elastic energy on precipitate morphologies is evident at $\sim$50 nm, while in the latter case, it only starts to become noticeable at $\sim$650 nm. There are three sources of error in the parameterization: the $C_{ijkl}^{\gamma}$ and $C_{ijkl}^{\gamma'}$ values, the interfacial energy, and the misfit strain.  To investigate the effect of variation in $C_{ijkl}$ values at a given misfit strain, we calculate $\overline{g_{el}}$ for two systems: one for a misfit strain of 0.53\% with $C_{ijkl}$ values calculated for 1173 K, and the other for a misfit strain of 0.1\% with $C_{ijkl}$ values calculated for 300 K, (i.e., substituting one set of $C_{ijkl}$ values for the other at each misfit strain).  Examining the values shown in Table \ref{tab:elastic-energy}, it is clear that the misfit strain has a much more significant effect on the elastic energy of a precipitate than any reasonably expected uncertainty in the exact $C_{ijkl}$ values.

\begin{table}
\centering
\caption{\label{tab:elastic-energy}The value of $\overline{g_{el}}$ as a function of $C_{ijkl}^{\gamma}$, $C_{ijkl}^{\gamma'}$ calculated at 300 K and 1173 K, and $\epsilon^T$ values.} 
\begin{tabular}{c c c c}
\hline
Temperature & $\epsilon^T=0.53\%$ & $\epsilon^T=0.1\%$ \rule{0pt}{2.6ex} \rule[-0.9ex]{0pt}{0pt}
\\\hline 
300 K & $2.80 \times 10^6 \textrm{ J/m}^3$ & $9.95 \times 10^4 \textrm{ J/m}^3$ \\ [5pt]
1173 K & $2.66 \times 10^6 \textrm{ J/m}^3$ & $9.45 \times 10^4 \textrm{ J/m}^3$ \\ [5pt]
\hline
\end{tabular}
\end{table}

The ratio of the elastic strain energy to the interfacial energy controls precipitate morphology, meaning that a smaller misfit strain paired with a smaller interfacial energy could yield similar microstructures.  The interfacial energy that we calculate by the density functional theory approach is significantly larger than that calculated from coarsening data (19 $\textrm{mJ/m}^2$ and 10 $\textrm{mJ/m}^2$ at 1073 K and 1173 K, respectively) \cite{meher2013coarsening}. In addition, there is significant scatter in high-temperature misfit strain results, which may be a result of differences in measurement technique \cite{yan2014alloying} or by approaching the $\gamma '$ solvus, which can shift somewhat depending on the exact composition \cite{yan2014alloying}.  However, the similarity of the experimental microstructures to the precipitate morphologies simulated with the 300 K parameter set indicate that the Co-Al-W $\gamma$/$\gamma'$ system has a $\overline{g_{el}}$/$\Gamma$ ratio of $3 \times 10^7 \ \textrm{m}^{-1}$. We argue that it may be possible to use this information to reduce uncertainty in the interfacial energy value. Careful measurement of the $\gamma/\gamma'$ misfit strain \textit{at the temperature for which the microstructure was evolved} combined with microscopy to characterize the exact length scales of the spherical to cuboidal transition could provide a new means of estimating the interfacial energy.  

\subsection{Multiple precipitates
\label{sub:multi-precip}}

We also perform multi-precipitate simulations to study how the microstructure is affected by precipitate interactions, both when the system is not externally stressed and when it is experiencing uniaxial tensile or compressive stress.  Given the results in Section \ref{sub:single-precip}, we choose the 300 K parameter set for the misfit strain and $C_{ijkl}$ values. We utilize mirror symmetry to simulate one-quarter, one-half, and one whole precipitate to study arrays of two, four, and eight precipitates.  In addition to reducing computational requirements, mirror symmetry allows us to examine equilibrium shapes of precipitate arrays without the confounding effect of particle coarsening that occurs when precipitates are different sizes;  we anticipate that coarsening occurs at much longer time scales than shape evolution driven by elastic and interfacial energies. Precipitates are initially spherical with a diameter of 200 nm, have a separation of 50 nm, and are aligned in the $\langle 100 \rangle$ directions. When applying an external stress, we study both when it is oriented normal to and parallel to the line of two precipitates or the plane of four precipitates (these two orientations are equivalent for eight precipitates). Precipitate evolution proceeds through two stages, similar to that observed in Ref.\ \cite{su1996dynamics}: a relatively rapid adjustment of precipitate shape in response to interfacial energy and the elastic energy of the precipitate, and a much slower motion of the centers of mass in response to the elastic stress fields induced by the other precipitates.

\begin{figure}
\centering
\subfloat[\label{fig:8part-0MPa}]{\includegraphics[scale=1.75]{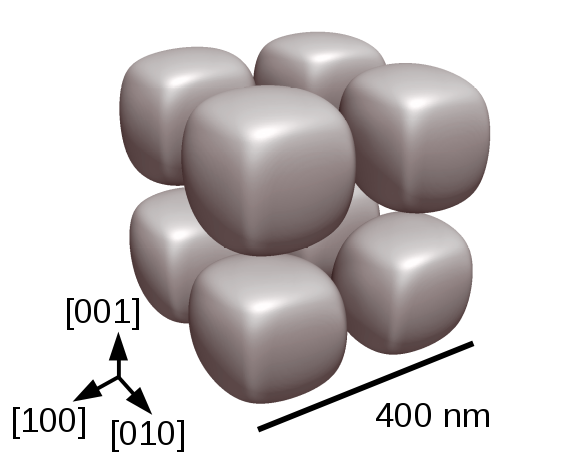}}
\subfloat[\label{fig:8part-p400MPa}]{\includegraphics[scale=1.75]{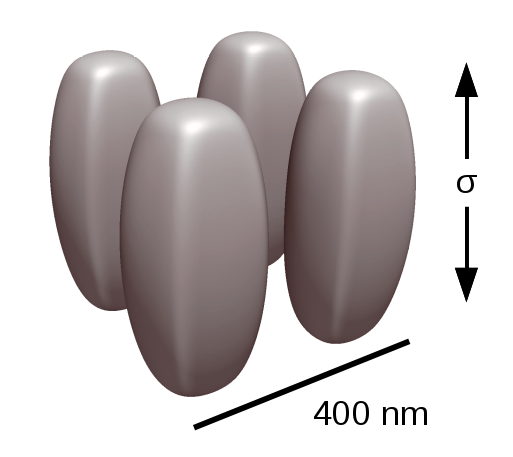}}
\subfloat[\label{fig:8part-m400MPa}]{\includegraphics[scale=1.75]{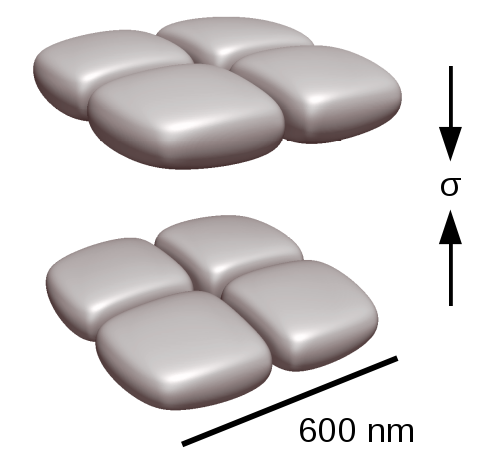}}

\caption{Characteristic microstructures develop as a function of the applied stress (stress orientation is indicated by the arrows).  a) No applied stress results in cuboidal precipitates,  b) 400 MPa applied in tension results in rods aligned parallel to the stress, c) 400 MPa applied in compression results in plates aligned perpendicular to the stress.   \label{fig:characteristic-structures}}
\end{figure}

We find that characteristic microstructures develop depending on the sign of the applied stress, which are similar to the morphologies found in the single-precipitate simulations (Fig.\ \ref{fig:characteristic-structures}). Without applied stress, precipitates remain approximately cuboidal and separated by an equilibrium channel width.  The precipitates in the arrays of four and eight are cuboidal with a size of 220 nm and 206 nm, respectively, and the channel widths  are 56 and 47 nm, respectively. The vertices of the precipitates at the centers of the arrays are more rounded than those at the exteriors of the arrays, similar to experimentally observed morphologies of isolated $\gamma'$ arrays in superalloys (e.g., Ni-Si $\gamma'$ arrays in Ref.\ \cite{doi1984effect}).  Conversely, the precipitates in the two-particle array become plate-like with their short axes parallel to the alignment direction of the precipitates.  These precipitates have long axes of 266 nm and short axes of 220 nm and are separated by a 69 nm channel.  The channel width appears to be a function of the size of the individual precipitates but not the total volume of the precipitates in the array, as the channel width increases as the individual precipitate size increases.  Note that channel widths for two arrays with the same number of precipitates but with different precipitate sizes are not compared here, which would provide additional information regarding the effect of array symmetry and precipitate size on channel width.

With an applied uniaxial stress of 400 MPa, the microstructure develops two length scales, which is characterized by arrays of large rods or plates composed of multiple rodlets or platelets (Figs.\ \ref{fig:8part-p400MPa} and \ref{fig:8part-m400MPa}).  
Similar to the single precipitates, rods form parallel to the applied tension, and plates form perpendicular to the applied compression. The centers of mass of the rods or plates move significantly to attain the equilibrium separation distance. Conversely, the equilibrium separation of the rodlets or platelets within the rods or plates is small, such that the rodlets or platelets are separated by narrow channels of $\gamma$.  For the precipitates in compression, the plate thickness is in the range of 130 nm, while the total plate width is on the order of 400 nm.  The plate separation is 190 -- 330 nm, with larger total array volumes corresponding with larger separations.  The $\gamma$ channel width separating platelets is 20 -- 30 nm, and larger individual platelets correspond with larger channel widths.  For precipitates under tension, the rod length is in the range of 420 -- 730 nm and the rod width is in the range of 175 -- 190 nm, with the rods separated by 84 -- 90 nm.  However, the rodlets within the rods neck and merge into one particle when a diffuse interface width of 10 nm is simulated, indicating that the channel spacing between the rodlets is less than 20 nm.  To investigate this further, we utilize an interface width of 5 nm, which is on the order of the physical interface width, and examine two two-precipitate simulations in which the applied stress is parallel to the precipitate alignment.  The rodlets within the rod are separated by 11 and 13 nm (the rodlets are 278 long by 148 nm wide, and 387 nm long and 180 nm wide, respectively).  Four- and eight-particle configurations are not studied because of the high computational cost and likelihood of little additional useful information being gathered.

\begin{figure}
\centering
\includegraphics[scale=1.75]{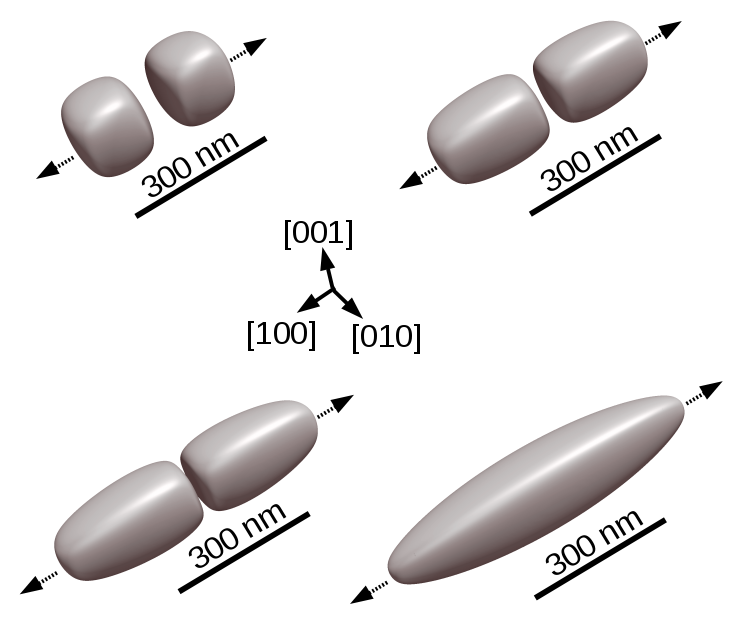}

\caption{Precipitates elongate and the width of the inter-precipitate $\gamma$ channel decreases with increasing applied tension, ultimately leading to precipitate merging when the stress is large enough. Upper left) 0 MPa, upper right) 200 MPa, lower left) 400 MPa, lower right) 800 MPa (direction indicated with dashed arrows). The interface width is 5 nm, similar to the experimentally observed interface width. \label{fig:2precip-tension}}
\end{figure}

We further investigate the effect of applied tension on precipitate morphology, because the rodlet merging observed for the unphysically wide diffuse interface indicates that precipitate merging is energetically favored.  We perform additional two-precipitate simulations with a diffuse interface width of 5 nm in which 200 MPa and 800 MPa are applied parallel to the direction of precipitate alignment. The precipitate volumes are within 20\%, reducing the confounding effect of precipitate volume on channel width.  The results are shown in Fig.\ \ref{fig:2precip-tension}, clearly indicating increasing precipitate aspect ratios and decreasing $\gamma$ channel width with increasing applied stress.  The $\gamma$ channel width between rodlets is 43 nm at no applied stress, 25 nm at 200 MPa, 11 nm at 400 MPa, and 0 nm (the rodlets neck and merge) at 800 MPa.  

\begin{figure}
\centering
\includegraphics[scale=2]{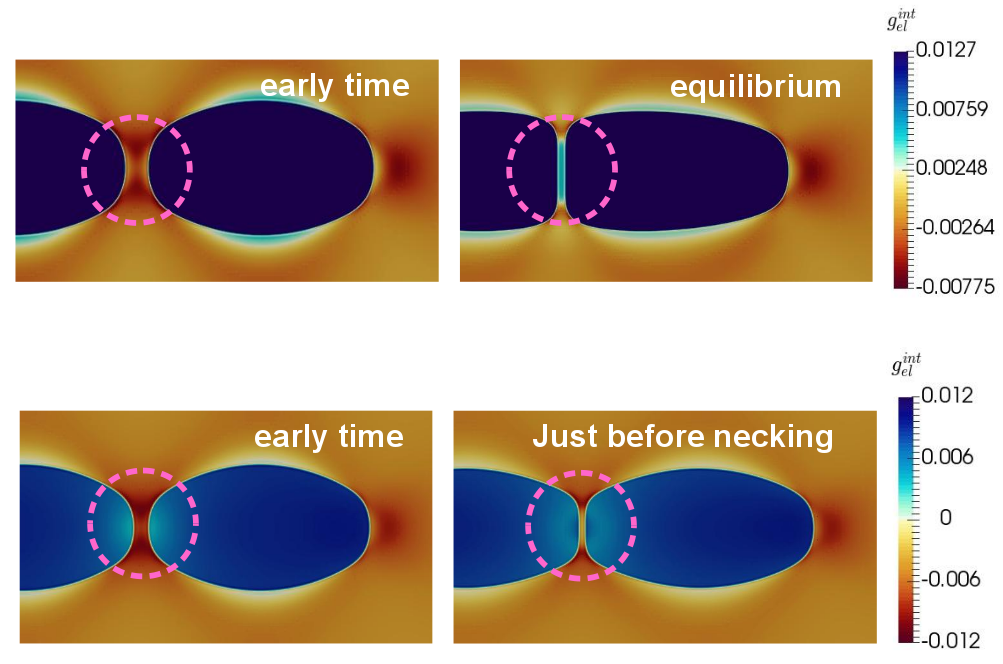}

\caption{The elastic interaction energy of the local stress field and the misfit strain of the precipitate affects the width of the $\gamma$ channel separating $\gamma'$ precipitates. When $g_{el}^{int}$ is negative, there is a driving force to bring the precipitates closer together; a positive energy impedes approach.  The precipitates are blue and the $\gamma$ channel separating them is indicated by the dashed circle.  Top) Precipitates with 400 MPa of tension applied parallel to the line of the precipitates.  At an early time in the precipitate evolution, the energy in the channel is negative; at equilibrium, the energy is positive.  Bottom) Precipitates with 800 MPa of tension applied. At the early time, the energy is negative and larger in magnitude than the 400 MPa case.  Just before the precipitates neck, the energy remains significantly negative, indicating a thermodynamic driving force for necking and merging. \label{fig:gel-int}}
\end{figure}

To understand the thermodynamic basis governing $\gamma$ channel width and whether precipitates merge or not, we examine the elastic interaction energy density, $g_{el}^{int} = \sigma_{ij}\epsilon^T_{ij}$ \cite{su1996dynamics,shen2006effect}, within the $\gamma$ channels. The elastic interaction energy indicates how the misfit strain of a precipitate will interact with the local stress field. When $g_{el}^{int}$ is negative, there is a driving force to bring the precipitates closer together; a positive energy impedes approach. An example is shown in Fig.\ \ref{fig:gel-int} for two precipitates with either 400 MPa or 800 MPa applied in tension. At the start of the simulations, the energy is negative and the precipitates become closer together. As evolution continues, the energy becomes positive in the 0 MPa, +200 MPa, +400 MPa, and -400 MPa cases, indicating a repulsive force opposing the precipitates from approaching more closely.  However, in the case of +800 MPa, the energy remains large and negative just before the precipitates neck. These results indicate that there is a thermodynamic basis governing precipitate separation and that there is a threshold stress to precipitate necking and merging while under applied tension (this may also be the case under compression).  Using a linear interpolation of $\gamma$ channel width and the value of applied stress, we predict that the channel width will be less than 5 nm with 470 MPa applied stress, which should lead to precipitate necking. 

We make several final notes.  Given the above results, the diffuse interface width is an important parameter whose effect on the results must be understood when performing quantitative studies (investigated, for example, in Ref.\ \cite{zhu2004three} with respect to coarsening kinetics).  We observe that particle necking and merging may occur when the diffuse interfaces of two precipitates start to overlap, but that the threshold stress at which necking and merging occurs is affected by the choice of interface width within the simulation. We find that the threshold stress decreases with increasing diffuse interface width, indicating that an accurate assessment of the threshold stress necessitates the use of the more computationally expensive physical interface width instead of a less expensive, unphysically large width. In addition, antiphase domain boundaries (APBs) can affect the particle merging process if the energy of the APB is sufficiently high \cite{wang1998field}.  The equilibrium channel widths that we find will not be affected by a difference in order parameter between the two particles, since the channel widths are solely a result of a repulsive elastic interaction energy. However,  at sufficiently high stress, the merging process can be affected by APB energy. This can affect both the precipitate shape evolution and the transformation kinetics in coarsening/rafting simulations. Furthermore, we note that we compare the simulated morphologies of isolated precipitates or precipitate clusters at small $\gamma'$ volume fractions with experimental morphologies at large $\gamma'$ volume fractions. We cannot quantitatively compare inter-precipitate spacings, because long-range elastic interactions between multiple precipitates at high volume fractions may affect the result. Both simulations with high precipitate volume fractions and experiments with low $\gamma'$ volume fractions would facilitate further analysis. Finally, we point out that when imaging microstructures, especially rafted structures, views should taken both parallel and perpendicular to the applied stress, as rods and plates can look the same depending on the viewing angle.
%
%

\section{Conclusion}
\label{Conclusion}

We examine the the thermodynamic forces that govern the equilibrium shapes of $\gamma'$  precipitates in the Co-Al-W system and precipitate coarsening behavior. Our work shows how the equilibrium shape is determined by the interplay of the interfacial energy and the elastic energy, the latter of which depends on the $\gamma$/$\gamma'$ misfit strain and the elastic stiffness of each phase.  In order to determine or at least bracket poorly known materials parameters, we make two sets of best guesses for materials parameters at 300~K and 1173~K, enabling us to determine how variation in the parameterization affects the results.  We examine individual precipitates and symmetrical arrays of multiple precipitates, and we simulate equilibrium morphologies with no applied stress, tensile stress, and compressive stress.  We also examine how precipitate morphology changes as a function of size for individual precipitates.

We make several conclusions regarding misfit strain and elastic energy in the Co-Al-W system.  We observe that the precipitate shapes simulated when using the 1173 K parameter set are not similar to experimental microstructures ($\epsilon^T=0.1\%$), but they are when simulated with the 300 K parameter set ($\epsilon^T=0.53\%$). This result indicates that the system has an elastic to interfacial energy ratio of $3 \times 10^7 \ \textrm{m}^{-1}$.  In addition, we find that uncertainty in the elastic strain energy density of a precipitate is primarily affected by the misfit strain as opposed to the elastic stiffnesses, meaning that it is important to characterize the misfit strain at each temperature the microstructures are evolved.  Furthermore, the system exhibits bifurcation behavior (elastic shape instability at a critical size). The first critical size is in between 180 -- 240 nm, above which the equilibrium shape of precipitates is plate-like with very low axial anisotropy ratios. A second critical size exists within 277 -- 319 nm. Over this size, precipitates exhibit a rod-like shape, with the ratio of the long:short axes increasing with increasing precipitate volume.  We suggest that this behavior explains the occurrence of the experimentally observed directional coarsening without the application of stress.  

Furthermore, we observe characteristic structures depending on the applied stress; these characteristic structures occur for both isolated precipitates and multi-precipitate arrays.  Precipitates become rod-like parallel to applied tension, and plate-like perpendicular to applied compression. Two length scales exist, in which large rods or plates are composed of multiple rodlets or platelets separated by narrow channels of $\gamma$ phase. These channels are thermodynamically favored until a threshold stress is reached.  We predict that, in tension, a stress of approximately 470 MPa is necessary to cause precipitate necking and merging via the overlap of the precipitates' diffuse interfaces.  We also find that accurate prediction of the threshold stress is affected by the choice of the diffuse interface width; the threshold stress is less than 400 MPa when the unphysically large width of 10 nm is chosen to reduce computational cost.  While mesh adaptivity greatly reduces the computational cost of the 3D simulations, they remain computationally intensive.  Unphysically wide diffuse interfaces may be used in most cases, but quantitative studies must always consider their effect on results. 

Our model may be used to rapidly explore the effect of interfacial and elastic energy on equilibrium microstructures.  Specifically, variations in misfit strain, elastic stiffness, and precipitate sizes may influenced by materials chemistry, such that this model may be helpful in predicting coarsened and rafted microstructures of possible new alloys. The model is quick to implement and gives useful microstructural information without including chemical free energy and solute diffusion, which can become very complex for superalloy compositions.  In addition, the model may easily be modified to include compositional information. With the addition of chemical diffusion, the kinetics of the microstructure evolution may be studied. For instance, the morphologies of evolving precipitates may be compared with equilibrium morphologies at the same size, allowing an understanding of the influence of diffusion kinetics on morphological evolution. 

\section*{Acknowledgments}
\label{Acknowledgments}
The work by A. M. J., O. G. H., P. W. V., S. S. N., and C. W. was performed under financial assistance award 70NANB14H012 from U.S. Department of Commerce, National Institute of Standards and Technology as part of the Center for Hierarchical Material Design (CHiMaD). A. M. J. and O. G. H. gratefully acknowledge the computing resources provided on Blues and Fission, high-performance computing clusters operated by the Laboratory Computing Resource Center at Argonne National Laboratory and the High Performance Computing Center at Idaho  National Laboratory, respectively. S. S. N. and C. W. performed computations using Quest High Performance Computing Cluster at Northwestern University and resources of the National Energy Research Scientific Computing (NERSC) Center, a DOE Office of Science User Facility supported by the Office of Science of the U.S. Department of Energy under Contract No. DE-AC02-05CH11231.  Results shown in this work are derived from work performed at Argonne National Laboratory. Argonne is operated by UChicago Argonne, LLC, for the U.S. Department of Energy under contract DE-AC02-06CH11357. Finally, we thank D. Seidman, D. Dunand, D. Sauza, J. Coakley, and E. Lass for their expertise and productive discussions regarding superalloys.


\bibliographystyle{Cosuperalloy}
\bibliography{Cosuperalloy.bib}


\end{document}